\documentclass[fleqn,12pt,twoside]{article}
\usepackage{espcrc1}

\usepackage{graphicx}
\usepackage[figuresright]{rotating}

\newcommand{\AmS}{{\protect\the\textfont2
  A\kern-.1667em\lower.5ex\hbox{M}\kern-.125emS}}

\newcommand*{\no}{\noindent}
\newcommand*{\bea}{\begin{eqnarray}}
\newcommand*{\eea}{\end{eqnarray}}
\newcommand*{\be}{\bea}
\newcommand*{\ee}{\eea}
\newcommand*{\pd}{\partial}
\newcommand*{\pdm}{\pd_{\mu}}

\newcommand*{\pref}[1]{(\ref{#1})}

\newcommand*{\nn}{\nonumber}

\title{Instantons, monopoles, vortices, and the Faddeev-Popov operator eigenspectrum}

\author{Axel Maas\address[IFSC]{Instituto de F\'isica de S\~ao Carlos, University of S\~ao Paulo, C.\ P.\ 369, 13560-970 S\~ao Carlos, SP, Brazil}}
       
\begin{document}

\maketitle

\begin{abstract}
The relation of confinement scenarios based on topological configurations and the Gribov-Zwanziger scenario is examined. To this end the eigenspectrum of the central operator in the Gribov-Zwanziger scenario, the Faddeev-Popov operator, is studied in topological field configurations. It is found that instantons, monopoles, and vortices contribute to the spectrum in a qualitatively similar way. Especially, all give rise to additional zero-modes and thus can contribute to the Gribov-Zwanziger confinement mechanism. Hence a close relation between these confinement scenarios likely exists.
\end{abstract}

\section{Introduction}

Several confinement scenarios exist today. These are able to explain various aspects of confinement, like the static quark-anti-quark potential or the absence of gluons from the physical spectrum. The scenarios are based on quite different mechanisms. Some scenarios rely on topological configurations, like center vortices or monopoles \cite{Greensite:2003bk}. On the other hand, e.\ g., the geometry of field configuration space plays a central role in the Gribov-Zwanziger scenario \cite{Zwanziger:1993dh,Gribov}.

However, all correct scenarios have necessarily to be equivalent. With an increasing understanding of the confinement mechanism from various view-points it thus becomes desirable to find the relation between these scenarios. First steps in this direction start to relate, e.\ g., the vortex and the dual superconductor picture \cite{Greensite:2003bk,deForcrand:2000pg}.

Here, a more distant relation is investigated: The one of the Gribov-Zwanziger scenario and of those involving topological configurations. The central element in the Gribov-Zwanziger scenario is the Faddeev-Popov operator, in particular its eigenspectrum. It is given in Euclidean space as
\be
-\pdm\left(\pdm+gf^{abc}A_\mu^c\right).\label{fp}
\ee
\no Here $g$ is the gauge coupling, $f^{abc}$ the structure constants of the gauge group (taken to be SU(2)), and $A_\mu^c$ the gauge field. $A_\mu^c$ will be selected to be the field of an instanton \cite{Bohm:2001yx}, an Abelian projected monopole \cite{Ripka:2003vv}, and a thick center vortex \cite{Diakonov:1999gg}, respectively.

It is worthwhile to note that by writing down $A_\mu^c$, the problem of determining the eigenspectrum of the operator \pref{fp} is complete \cite{Maas:2006nq}. It is not necessary to specify a gauge. In fact, the results are valid in any gauge which is satisfied by the field configuration $A_\mu^c$. However, the operator \pref{fp} may have different significance for the various gauges satisfied by the field configuration $A_\mu^c$. Thus, in general one should investigate generalized Faddeev-Popov operators for general gauges\footnote{Thanks to Attilio Cucchieri for pointing this out.}.

Here only the operator \pref{fp} will be discussed. It is the relevant operator in Landau gauge. In case of a (spatially-aligned) vortex or a monopole the results do not change when replacing the operator with the one relevant in Coulomb gauge. The operator differs only by restricting to the spatial subspace, and these field configurations do not depend on time.

\section{Eigenspectrum}

To obtain the eigenspectrum, the eigenequation
\be
-\pdm\left(\pdm+gf^{abc}A_\mu^c\right)\phi^b=\omega^2\phi^a\nn
\ee
\no has to be solved. In this context only the eigenvalues are interesting, as the Gribov-Zwanziger scenario predicts an enhancement of the spectrum at or near zero eigenvalue. Various investigations using lattice gauge theory have shown that such an enhancement indeed exists \cite{Cucchieri:2006tf,Greensite:2004ur}. Furthermore, it has been found that this enhancement is sensitive at least to vortices: Removal of thin center vortices in lattice calculations also removes the enhancement \cite{Greensite:2004ur}.

To evade the finite volume effects in lattice calculations, a complementary ansatz is an analytical continuum investigation \cite{Maas:2006nq,Maas:2005qt}. By specifying the field configuration completely, e.\ g.\ a single instanton, the eigenvalue problem is well-posed and can be solved. However, in this case also the eigenfunctions have to be determined to make sure that an obtained eigenvalue is admissible: In general, the eigenfunctions can exhibit a behavior which is more singular or diverging as the gauge-filed configuration and thus are not permitted.

The problem is then mathematically equivalent to solving a static Schr\"odinger equation in 4+1 dimensions. The field configurations turn out to lie all in the first Gribov region or on its boundary \cite{Maas:2005qt}. Hence, the system does not permit bound states, but possibly zero-energy states. Therefore it corresponds to a scattering problem, and the eigenstates cannot be expected to be localized or normalizable. This is already clear in the vacuum, where $A_\mu^c$ vanishes. Then the operator is the (negative) Laplacian, and the solutions are free waves. In fact this non-normalizability plays an important role in the Gribov-Zwanziger scenario \cite{Zwanziger:1993dh}. It is thus comforting that this property emerges naturally.

In case of an instanton or a vortex, it is found that 3 and 2(flux-1) non-trivial zero modes exist, respectively \cite{Maas:2005qt}. The positive, continuous eigenspectrum is qualitatively unaltered by the presence of these configurations. In both cases the solutions can be found by separating the angular variables and decoupling the internal color index. The remaining radial part can then be solved by a series ansatz.

The situation is a little bit different in case of an Abelian-projected monopole. Given its field configuration in spherical coordinates \cite{Ripka:2003vv}
\be
A_\mu^a=-\delta^{3a}\frac{1}{g}\vec{e}_\varphi\frac{1+\cos\theta}{r\sin\theta}
\ee
\no it follows immediately that color 3 decouples and becomes trivial. This is like in the vortex case \cite{Maas:2006nq,Maas:2005qt}. Thus only the color-off-diagonal components can produce a non-trivial behavior.

Contrary to the case of the instanton and the vortex, the non-trivial part of the field configuration resides in the angular part. Consequently a solution can be constructed by separating the coordinates $t$ (as plane waves $\exp(ikt)$), $r$ (as Bessel functions $J_n\left(\sqrt{\omega^2-k^2}r\right)$), and $\phi$ (as $\exp(im\phi)$). Combining $\phi^1$ and $\phi^2$, it is possible to reduce the system to one ordinary differential equation,
\be
\left(\kappa+\frac{1}{\sin\theta}\pd_\theta\sin\theta\pd_\theta-\frac{(m^2+m(1+\cos\theta))}{\sin^2\theta}\right)\phi^+=0\label{deq}
\ee
\no where $\kappa$ is a constant depending on $n$ and $\phi^+=\phi^1+i\phi^{2}$. The same equation with $m\to -m$ holds for $\phi^1-i\phi^{2}$, similar to the vortex case \cite{Maas:2005qt}. Transforming to the variable $x=1+\cos\theta$, it is possible to solve the equation with the series
\bea
\phi_{\kappa m}^+(x)&=&x^{\frac{|m|}{2}}\sum_{n\ge 0} a_n x^n\label{series}\\
a_n&=&((-4(n-2)(n-1)+4\kappa-2(2(n-1)-1)|m|-m^2)a_{n-2}\nn\\
&&+4((2n-1)(2n-2)-2\kappa+(1+4(n-1))|m|+m+m^2)a_{n-1})\frac{1}{16n(n+|m|)}.\nn
\eea
\no $a_0$ is a free normalization constant and $a_{-1}$ vanishes. The second solution to \pref{deq} is too singular, and is thus dismissed.

Note that this equation is only indirectly dependent on the eigenvalue $\omega$: If $\omega$ is non-zero, any value of $n$ is admissible. For $\omega=0$ only $n=0$ is possible, yielding $\kappa=-1/4$. To count the zero-modes it is thus necessary to check which of the (integer) values of $m$ lead to admissible solutions.

Investigating the limit $x\to 2$ in \pref{deq}, and thus towards the field singularity, it is found that the solution behaves as $(x-2)^{\pm\sqrt{2m+m^2}/2}$. Only one of these solution corresponds to \pref{series}. Which of both can practically only be decided numerically, as the series is hypergeometric. This is as in the case of the vortex \cite{Maas:2005qt}.

It turns out that there is only one solution\footnote{The value $m=0$ leads to a trivial, constant zero-mode.} at $m=-2$ and one at $m=-1$. The former behaves as $\ln(x-2)$, while the other is a linear combination of $\sin(\sqrt{(-m(2+m)}\ln(x-2)/2)$ and $\cos(\sqrt{(-m(2+m)}\ln(x-2)/2)$. Thus there are two non-trivial zero-modes for $\phi^+$ and two more for the solution $\phi^-$, altogether four.

Hence also the monopole provides a finite number of zero-modes, similar to an instanton or a vortex. Thus also monopoles can be relevant in the Gribov-Zwanziger confinement scenario. This also implies that all types of topological configurations investigated so far are qualitatively equally relevant.

In addition, the color structure is quite interesting. While the Cartan component of the zero-modes is trivial, the off-diagonal elements are highly non-trivial. These off-diagonal elements can couple to the Cartan gluons in Abelian gauges. Hence it seems possible that there might exist a mechanism how such non-trivial off-diagonal elements can help in confining the Cartan gluon in the dual superconductor mechanism. That would solve this long-standing problem in such models \cite{Ripka:2003vv}. However, this is at this stage pure speculation, but still an interesting direction.

\section{Conclusions}

With increasing and better understanding of the confinement mechanism in various scenarios, their unification has become an important task. Only if it is possible to connect all those different facets it will possible to speak of a real understanding of confinement.

The various results found in lattice calculations \cite{Greensite:2004ur} and those presented here and elsewhere \cite{Maas:2005qt} show that a direct connection exists between topological configurations and the Gribov-Zwanziger scenario. Moreover, this implies that the field configurations responsible for the static inter-quark potential and for the absence of gluons from the physical spectrum are possibly indeed the same. This would be very convenient. It meets the natural assumption that the confinement of quarks and gluons has the same origin. Nonetheless, these results on vortices, monopoles and instantons do not single out a particular relevant configuration, but all types seem to be involved. This coincides with results which point to a very intricate relation between these objects \cite{deForcrand:2000pg}.

Still, this is only a first step in such an unification process. Many more will be necessary.\\

\no{\bf Acknowledgment}

The author thanks the organizers for this interesting meeting and the opportunity to present this work. This work was supported by the DFG under grant MA 3935/1-1.


\begin{thebibliography}{9}

\bibitem{Greensite:2003bk}
J.~Greensite,
Prog.\ Part.\ Nucl.\ Phys.\  {\bf 51} (2003) 1 and references therein.

\bibitem{Zwanziger:1993dh}
D.\ Zwanziger,
Nucl.\ Phys.\ B {\bf 412}, 657 (1994).

\bibitem{Gribov}
V.\ N.\ Gribov,
Nucl.\ Phys.\ B {\bf 139}, 1 (1978);
D. Zwanziger,
Phys.\ Rev.\ D {\bf 65}, 094039 (2002);
Phys.\ Rev.\ D {\bf 67}, 105001 (2003).
Phys.\ Rev.\ D {\bf 69}, 016002 (2004),
and references therein.

\bibitem{deForcrand:2000pg}
P.~de Forcrand and M.~Pepe,
Nucl.\ Phys.\ B {\bf 598} (2001) 557;
J.~Ambjorn, J.~Giedt and J.~Greensite,
JHEP {\bf 0002} (2000) 033;
P.~Y.~Boyko {\it et al.},
arXiv:hep-lat/0607003;
H.~Reinhardt,
[arXiv:hep-th/0112215].

\bibitem{Bohm:2001yx}
M.~Bohm, A.~Denner and H.~Joos,
\textit{Gauge theories of the strong and electroweak interaction}, (Teubner, Stuttgart, 2001) 784p.

\bibitem{Ripka:2003vv}
G.~Ripka,
arXiv:hep-ph/0310102.

\bibitem{Diakonov:1999gg}
D.~Diakonov,
Mod.\ Phys.\ Lett.\ A {\bf 14} (1999) 1725;
D.~Diakonov and M.~Maul,
Phys.\ Rev.\ D {\bf 66} (2002) 096004.

\bibitem{Maas:2006nq}
A.~Maas,
arXiv:hep-th/0603087.

\bibitem{Cucchieri:2006tf}
A.~Cucchieri, A.~Maas and T.~Mendes,
Phys.\ Rev.\ D {\bf 74} (2006) 014503.
A.~Sternbeck, E.~M.~Ilgenfritz and M.~Muller-Preussker,
Phys.\ Rev.\ D {\bf 73} (2006) 014502.

\bibitem{Greensite:2004ur}
J.~Greensite, {\v{S}}.~Olejn{\'{i}}k and D.~Zwanziger,
JHEP {\bf 0505} (2005) 070
[arXiv:hep-lat/0407032].

\bibitem{Maas:2005qt}
A.~Maas,
arXiv:hep-th/0511307, accepted by Eur.\ Phys.\ J.\ C.

\end{thebibliography}
\end{document}